\documentclass[fleqn,10pt]{wlscirep}
\usepackage{graphicx}
\usepackage{color}
\usepackage{diagbox}
\usepackage{multirow}
\usepackage{subfigure}

\title{Experimental realization of state transfer by quantum walks with two coins}

\author[1,2*]{Yun Shang}
\author[1,3**]{Meng Li}

\affil[1]{Institute of Mathematics, Academy of Mathematics and Systems Science, Chinese Academy of Sciences, Beijing 100190, China}
\affil[2]{NCMIS, MDIS, Academy of Mathematics and Systems Science, Chinese Academy of Sciences, Beijing, 100190, China}
\affil[3] {School of Mathematical Sciences, University of Chinese Academy of Sciences, Beijing 100049, China}

\affil[*]{ shangyun602@163.com}
\affil[**]{limeng162@mails.ucas.ac.cn}

\begin{abstract}
 Quantum state transfer is an important task for scalable quantum complex networks. By selecting quantum walks with two coins as the basic model and two coin spaces as the communication carriers, we successfully implement some perfect state transfer tasks on various graphs (EPL, \textbf{124} (2018) 60009) \cite{Shang_2019}. Here, we provide the first experimental realization of this scheme by IBM Quantum Experience platform. Especially, we demonstrate the transfer of Bell state, GHZ state and W state on complete graph through this quantum apparatus. Furthermore, we notice that our protocols have high fidelity by performing quantum state tomography.

\end{abstract}

\begin{document}

\flushbottom
\maketitle
\thispagestyle{empty}

\section{Introduction}

In order to set up scalable quantum networks, perfect state transfer (PST) between network sites becomes a crucial important problem.
It was Bose who first took spin chains as the quantum communication channel in the study of quantum computing \cite{Bose_2003}. After that, there appeared a fruitful research on spin chains transfer \cite{christandl_2004, Burgarth_2005, Gong_2007, Di_2008, Gualdi_2008, Petrosyan_2010, Nikolopoulos_2014}. Since state transfer is affected by disorder, coherent and nearest-neighbor coupling relation of particles, in some cases, it is very difficult to achieve perfect state transfer for spin chains \cite{Hu_2012, Li_2009, Abliz_2006, Nikolopoulos_2014}. Soon, it was found that continuous-time quantum walks can be applied to describe the time evolution of qubit state transfer on a spin chain \cite{christandl_2004}.
Then studying state transfer by various quantum walk models becomes an interesting topic\cite{Kendon_2011, Yalcinkaya_2015, Stefanak_2016, Nitsche_2016}.
Correspondingly, experimental realizations of perfect state transfer have been reported \cite{Zhang_2005, Bellec_2012, Perez-Leija_2013, Chapman_2016}.
Bellec et. al presented the construction of PST photonic lattice  \cite{Bellec_2012}.
A long-range coherent quantum transport in photonic lattices was given by Perez-Leija et. al \cite{Perez-Leija_2013}.
Chapman et. al implemented the transport of quantum entanglement in PST photonic lattice \cite{Chapman_2016}.
Since transport problems are also known as hitting problems in computer science,
the experimental realization of quantum fast hitting on Hexagonal graphs was presented by Tang et. al \cite{Tang_2018}.

Different from existing models, we proposed some new quantum communication protocols through the model of quantum walks controlled by two coins.  By the use of two coin operators alternatively, we can realize perfect state transfer for high dimensional states on more general graph structures \cite{Shang_2019}.

Recently, an open platform named IBM Quantum Experience, which provides access to all users through a cloud-based webpage, attracts many people's concerns.
It is a quantum computer based on superconducting flux qubits \cite{IBM_2017}.
In order to detect and explore more applications coming from quantum world, many researchers have designed, tested and performed their experiments on this platform during the past few years. Certainly, many theoretical protocols have also been detected and proved in this platform \cite{Alsina_2016, Devitt_2016, Behera_2017, Hebenstreit_2017, Vishnu_2018, Chatterjee_2019}.

In this paper, we first demonstrate experimental implementation of state transmission by quantum walk models using two coins on IBM Quantum Experience platform. According to theoretical scheme \cite{Shang_2019}, we encode the transferred information in the 1-st coin space, and then alternatively run two coins operators, by simple correcting operation, we can successfully obtain the transferred information at the target site.
Based on the above scheme, selecting qubit state, Bell state, GHZ state, and W state as examples, we design quantum circuits and run them on the IBM interface respectively, and successfully achieve their transfer on cycle or complete graph respectively.
This article is arranged as follows. In Section 2, we will briefly introduce the model of quantum walks on graph with many coins and the setup of our experiment. In Section 3, we give the transfer realization of a qubit state on cycle. Then we illustrate the accuracy of our protocol by performing quantum state tomography.
In Section 4, we mainly consider transfer realization on the complete graph. Section 4.1, 4.2 and 4.3 narrate the experimental implementation of transferring the single qubit state, Bell state, GHZ state and W state on the complete graph respectively. Then we give the conclusion in Section 5.

\section{Preliminaries}

For quantum walks with one coin on an arbitrary graph $G(V,E)$, it can be depicted by the iteration of an unitary operator $U$ that applies on a composite Hilbert space $\mathcal{H}=\mathcal{H}^{P}\otimes\mathcal{H}^{C}$. Here $\mathcal{H}^{P}$ is called a position space which is spanned by vertexes in $V$. $\mathcal{H}^{C}$ is called a coin space which is spanned by edges in $E$.
Each step of the quantum walk is performed by $U=S\cdot (I\otimes C)$, where $S$ is the conditional shift operator that applies on the combined space, and $C$ is the coin operator applying on the coin space.
The conditional shift operator is
\begin{equation}
S=\sum_{i,e}|k\rangle\langle i|\otimes|e\rangle\langle e|,
\end{equation}
where the label $e$ represents a directed edge from vertex $i$ to vertex $k$.
In addition, the basis states of the walker are $|v,i\rangle$, where $v\in V$, $i\in \{0, 1 ,\cdots, d-1\}$, and $d$ is the degree of of vertex $i$. Thus the state of the quantum walks can be described as 
\begin{equation}
|\varphi(t)\rangle=\sum_{v,i}a_{v,i}|v,i\rangle.
\end{equation}

It was Brun who first introduced the model of quantum walks driven by many coins \cite{Brun_2003}.
As for the quantum walk with $m$ coins, it consists of one walker space and $m$-coins space. The state of the whole system is denoted by a vector in the Hilbert space $\mathcal{H}=\mathcal{H}^{P}\otimes\mathcal{H}^{C_1}\otimes \cdots\otimes\mathcal{H}^{C_m}$. The unitary evolution that results from flipping the $k$th coin is given by $U_k=(S_{k}\otimes I)\cdot(I\otimes C_{k})$, where $S_k$ acts on the combination space of position and $k$-th coin space, and $C_{k}$ acts on the $k$-th coin space. If we take a total of $t$ steps, the state will evolve into
\begin{equation}
|\varphi(t)\rangle=(U_m\cdot\cdot\cdot U _1)^{t/m}|\varphi(0)\rangle.
\end{equation}
In this paper, we mainly discuss quantum walks driven by two coins.

Here, we mainly use ``$ibmq_{\_}5_{\_}yorktown-ibmqx2$" and ``$ibmq_{\_}vigo$" which are 5-qubits quantum computers. Also, we consider ``$ibmq_{\_}16_{\_}melbourne$" which is a 14-qubits quantum computer to handle more complex quantum circuits.
For simplicity, we will write them as ``$ibmqx2$", ``$ibmq_{\_}vigo$" and ``$ibmq_{\_}16$" respectively.
In order to provide the related experimental realization on these real quantum processors, we need to design the quantum circuit that accords with the connectivity of the corresponding quantum chips \cite{IBM_2017}.
We represent the $CNOT$ gate that $i$ is the control qubit and $j$ is the target qubit as $C_{i,j}$.
In this paper, we mainly use two ways to design the suitable quantum circuit performed on quantum chips:
\begin{equation}
C_{i,j}=(H\otimes H)C_{j,i}(H\otimes H), \ \ \ \ \ \ \ C_{i,k}=C_{i,j}C_{j,k}C_{i,j}C_{j,k}.
\end{equation}
In addition, we might use the implementation of the $Toffoli$ gate with elementary single qubit gate and $CNOT$ gate \cite{Nielsen_2002}.

Recently, we successfully implement generalized quantum teleportation and perfect state transfer protocols on various graphs by selecting quantum walks driven by two coins as the basic model and two coin spaces as the communication carriers \cite{Shang_2019,Wang_2017 }.
The experimental realizations of generalized quantum teleportation by IBM quantum platform have been given by Chatterjee et. al \cite{Chatterjee_2019}.
Here, we would give the experimental realizations of state transfer scheme using IBM quantum platform.
Most importantly, it is the first time to provide an experimental realization of state transfer on different graphs using quantum walk model driven by two coins.

\section{Transfer realization of a qubit state on cycle}

First, let us focus on a qubit state transfer on $l$-cycle (i.e., a cycle with $l$ vertices). Obviously, the degree of each vertex is 2 and the coin space is spanned by $|0\rangle,|1\rangle$. And the conditional shift operator acting on the combination space for position and coin space is described as:
\begin{equation}
$$\tilde{S}=\sum_{i=0}^{l-1}(|(i+1)\bmod\,l\rangle\langle i|\otimes|0\rangle\langle0|+|(i-1)\bmod\,l\rangle\langle i|\otimes|1\rangle\langle1|).$$
\end{equation}

Compared with the previous paper \cite{Yalcinkaya_2015}, we have already achieved perfect state transfer on cycles with any number of vertices (not just on $l$-cycle where $l$ is an even number) and from the start point to any target location (not only from the start location to the opposite location) by introducing one more coin space \cite{Shang_2019}.

\subsection{Transfer realization of a qubit state on 8-cycle}

\begin{figure}[htb]
\centering
\includegraphics[width=8cm]{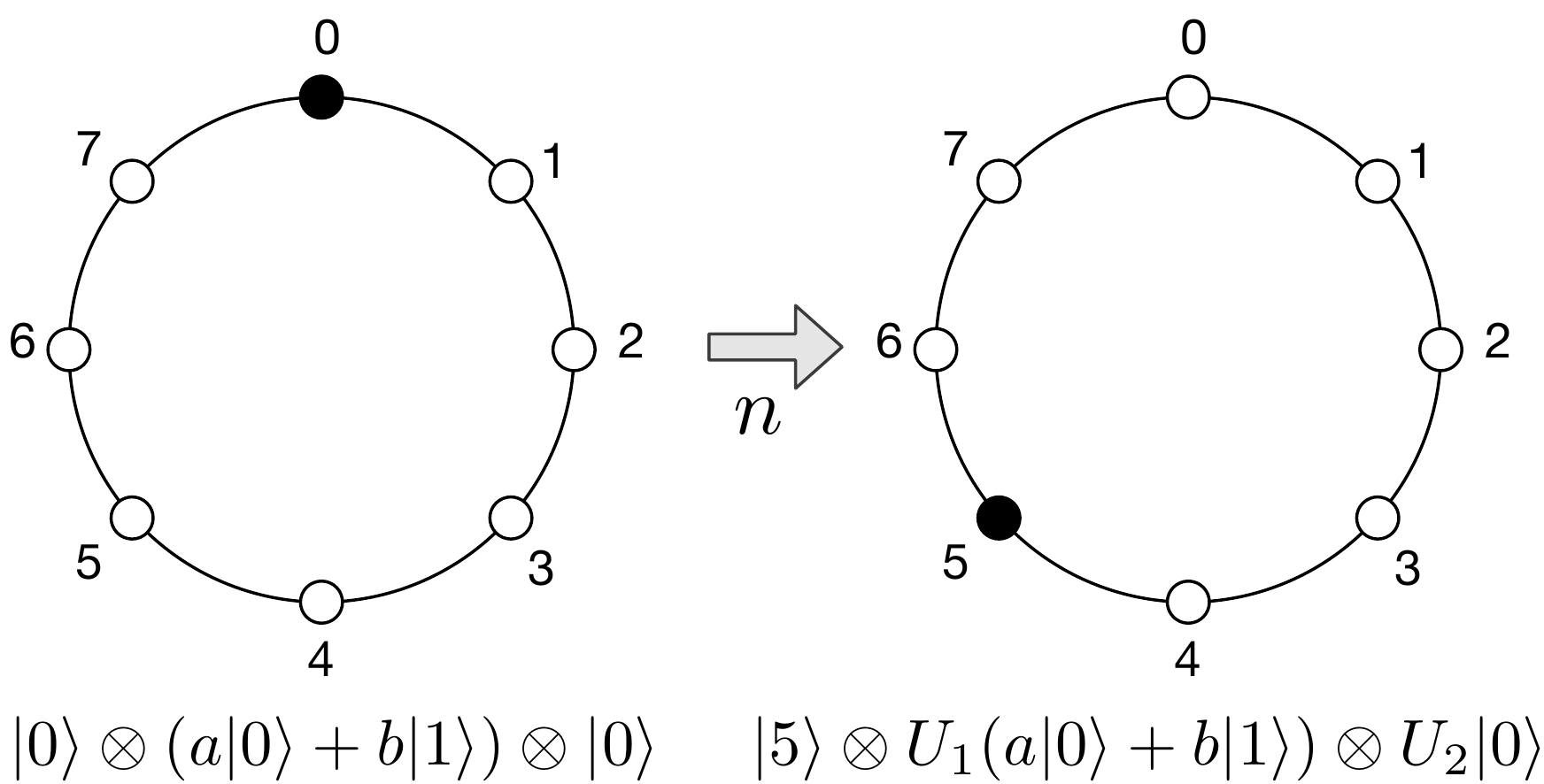}
\caption{Qubit state transfer on cycles. $U_{1}$ and $U_{2}$ are the recovery operator applying on the 1-st coin space and 2-nd coin space respectively.}
\label{fig:8-cycle}
\end{figure}

Here we take transferring a qubit state on a cycle with 8 vertices from $x=0$ to $x=5$ as an example, and give its experimental realization on IBM quantum experience platform. The total process can be described in Figure \ref{fig:8-cycle}.
According to our theoretical results \cite{Shang_2019}, we need to perform seven-step quantum walks by alternatively using the two coins totally. In the 1-st step and the 6-th step, we choose $C_{1}=X$ and $C_{2}=X$ respectively. For the rest of the steps, let coin operators be identity operators.
At last, we select Pauli operator $X$ as the correct operator $U_{1}$.

In order to understand its circuit diagram more clearly, we use the binary representation of these vertices. Hence the basis state $|000\rangle $, $|001\rangle$, $\cdots$, $|111\rangle$ represent $|0\rangle$, $|1\rangle$, $\cdots$, $|7\rangle$ respectively.
Without loss of generality, the state that we try to transfer could be assumed to be $\frac{1}{2}|0\rangle+\frac{\sqrt{3}}{2}|1\rangle$, which can be prepared by using the $U3(\frac{2\pi}{3},0,\frac{\pi}{2})$ gate on the IBM platform.
Note that $U3(\theta,\phi,\lambda)$ is a parameterized gate, which can be  described as follows:
\begin{equation}
U3(\theta,\phi,\lambda)=
\left(
  \begin{array}{cc}
    \cos(\frac{\theta}{2}) & -e^{i\lambda}\sin(\frac{\theta}{2}) \\
    e^{i\phi}\sin(\frac{\theta}{2}) & e^{i(\lambda+\phi)}\cos(\frac{\theta}{2})
  \end{array}
\right),
\end{equation}
Hence, the original state for the whole system is
\begin{equation}
|\phi\rangle^{0}=|000\rangle(\frac{1}{2}|0\rangle+\frac{\sqrt{3}}{2}|1\rangle)|0\rangle.
\end{equation}
And it is not difficult to infer that the final state is
\begin{equation}
|\phi\rangle^{7}=|101\rangle(\frac{1}{2}|1\rangle+\frac{\sqrt{3}}{2}|0\rangle)|1\rangle.
\end{equation}
which is just what we want to achieve with the help of recovery operator $U_{1}=X$.

\begin{figure}[htb]
\centering
\includegraphics[width=18cm]{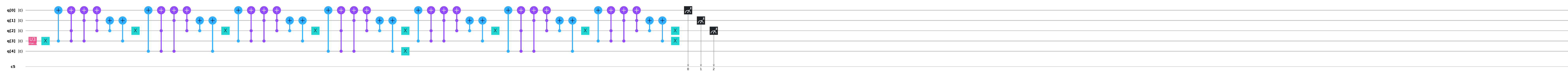}
\caption{Quantum circuit diagram of qubit state transfer on 8-cycle from $x=0$ to $x=5$. $q[0]$, $q[1]$, $q[2]$ stand for the position space. $q[3]$ and $q[4]$ stand for 1-st coin space and 2-nd coin space respectively. The measurements are in Z-basis for the position space. Here we use the $U3(\frac{2\pi}{3},0,\frac{\pi}{2})$ gate.}
\label{fig:8cycleposition}
\end{figure}

\begin{figure}[htb]
\centering
\includegraphics[width=18cm]{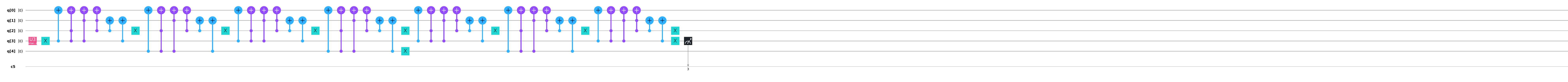}
\caption{Quantum circuit diagram of qubit state transfer on 8-cycle from $x=0$ to $x=5$. $q[0]$, $q[1]$, $q[2]$ stand for the position space. $q[3]$ and $q[4]$ stand for 1-st coin space and 2-nd coin space respectively. The measurements are in Z-basis for the 1-st coin space. Here we use the $U3(\frac{2\pi}{3},0,\frac{\pi}{2})$ gate.}
\label{fig:8cyclestate}
\end{figure}

\begin{figure}[htb]
 \centering
 \subfigure[]{\label{fig:8cyclepositionresult}
 \includegraphics[width=8cm]{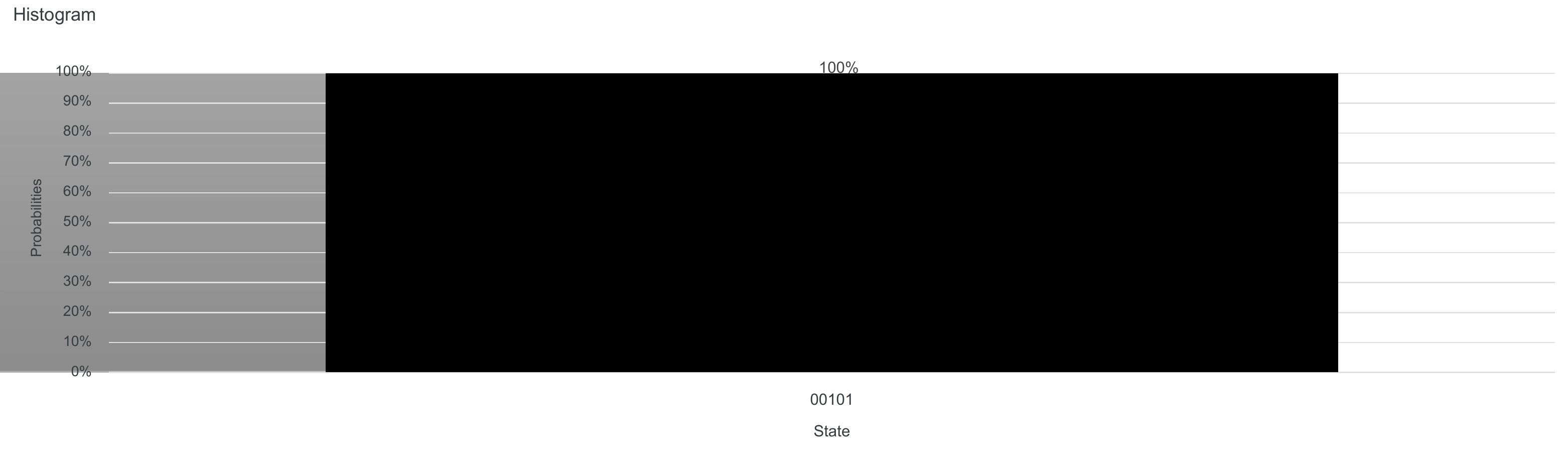}}
 \subfigure[]{\label{fig:8cyclestateresult}
 \includegraphics[width=8cm]{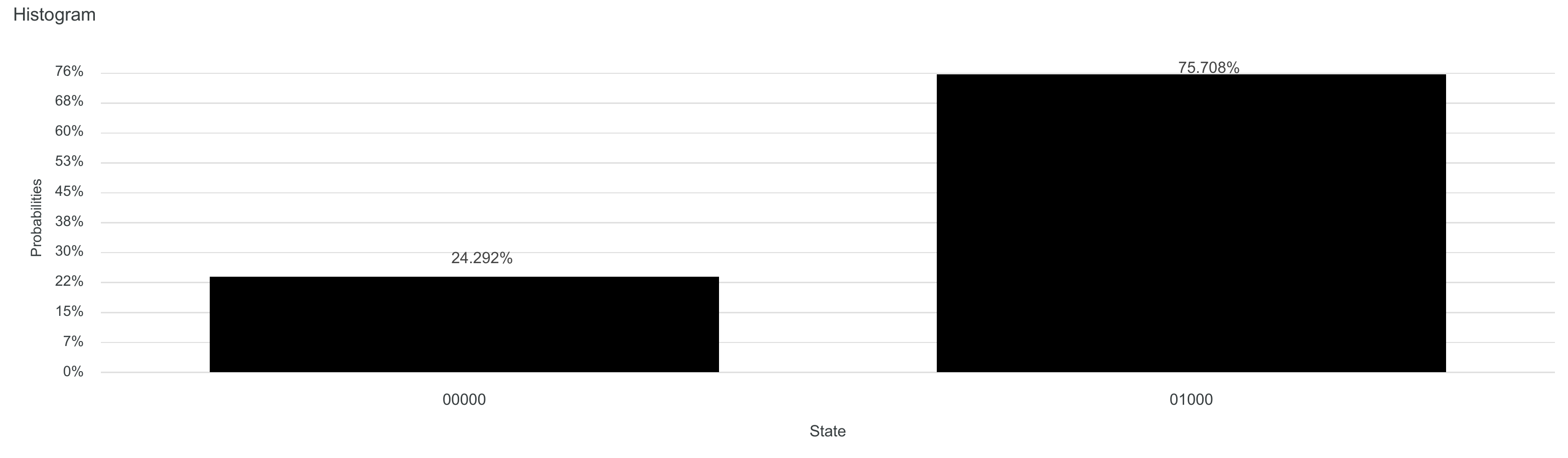}}
 \renewcommand{\figurename}{Figure}
 \caption{(a), (b) show the probability results of target position and transferred state after measurement respectively.}
 \label{fig:8cycleresult}
 \end{figure}

Based on the above discussion, we give the corresponding IBM quantum circuit diagrams which can be seen in Figure \ref{fig:8cycleposition} and Figure \ref{fig:8cyclestate}.
It is worth noting that every experiment has been run with 8192 number shots, in which the number of shots denote the number of times that measurements are performed.
And they were performed on the ``$simulator$" provided by IBM interface.
The results about the target position and transferred state are shown in Figure \ref{fig:8cycleresult}.
Comparing with the theoretical result (the probability at position $x=5$ is equal to 1; the probability of obtaining state $|0\rangle$ and $|1\rangle$ are $0.25$, $0.75$ respectively), the simulated result (on ``$simulator$") exists minor error.

\subsection{Quantum state tomography}
Although the simulation result seems very close to the theoretical value, it would produce noise during the implementation of quantum circuit.
In order to see the quality of simulation, we now proceed to perform quantum state tomography \cite{Paris_2004}. It is a well known process of reconstructing an unknown quantum state $|\phi\rangle$. In this case, for a given state, we need to calculate the fidelity between its theoretical and experimental (simulation) density matrices to determine the success degree of this protocol. For the given state $|\phi\rangle$, its theoretical density matrix is denoted as
\begin{equation}
\rho^{T}=|\phi\rangle\langle\phi|.
\end{equation}
Its experimental density matrix (for single qubit) can be described by
\begin{equation}
\rho^{E}=\frac{1}{2}(<X>X+<Y>Y+<Z>Z+I),
\end{equation}
where $X$, $Y$ and $Z$ are Pauli matrices.
Due to the measurements performed on this platform are all in computational basis (Z-basis), we need to provide other measurement bases by executing appropriate gate operations.
For example, to obtain the value of $<Y>$, we need to perform measurement in $Y$-basis. In this case, we could execute the conjugate transpose of phase gate $S^{\dagger}$ and Hadamard gate $H$ in sequence before a measurement in Z-basis.

For our initial state $|\phi\rangle=\frac{1}{2}|0\rangle+\frac{\sqrt{3}}{2}|1\rangle$, we can obtain these three average values by running the quantum circuit 10 times on $simulator$ and for each run the number of shots is 8192 to see the quality of simulation on IBM interface. The detail about the probability result is shown in Appendix A.
According to the formula and data, we can obtain the theoretical and the experimental density matrices as followings£»
\begin{equation}
\rho^{T}=
\left(
  \begin{array}{cc}
    0.2500 & 0.4330 \\
    0.4330 & 0.7500
  \end{array}
\right),
\end{equation}
\begin{equation}
\rho^{E}=
\left(
  \begin{array}{cc}
    0.2479 & 0.4323-0.003i \\
    0.4323+0.003i & 0.7521
  \end{array}
\right).
\end{equation}
The fidelity between theoretical and experiment (simulate) density matrix could be calculated by $F(\rho^{T},\rho^{E})=(tr(\sqrt{\sqrt{\rho^{T}}\rho^{E}\sqrt{\rho^{T}}}))^{2}$, which is found to be approximately equal to 1.

\section{Transfer realization of a qudit state on the complete graph}

As is known to all, a qudit state can be written as $\sum_{k=0}^{n-1}c_k|k\rangle$, where $|c_k|^2=1$.
First, we introduce the outline to perfectly transmit a qudit state on an arbitrary complete graph  by using two coins quantum walk model \cite{Shang_2019}.
A complete graph with $n$ nodes is denoted as $n$-complete graph.
Every coin space is spanned by $|0\rangle,\cdots,|n-1\rangle$.
The conditional shift can be described as
\begin{equation}
\hat{S}=\sum_{i,j=0}^{n-1}|(i+j)\bmod\,n\rangle\langle i|\otimes|j\rangle\langle j|.
\end{equation}
The original state of this system could be written as
\begin{equation}
|\phi\rangle^{0}=|0\rangle\otimes(\sum_{i=0}^{n-1}c_i|i\rangle)\otimes|0\rangle.
\end{equation}

In order to transmit the initial state from $0$ to $y$ on an arbitrary $n$-complete graph ($y\in \{0, 1, \cdots, n-1\}$), we need to perform $2n$-step quantum walks by alternatively using the two coins according to the given scheme \cite{Shang_2019}. To be specific, we only choose $C_{2}=X_{n}$ in the $(2n-2y+2)$-th step where $X_{n}$ is the generalized Pauli X operator. For the rest steps, the coin operators are all identity operators.

Since the coin operator $C_{1}$ or $C_{2}$ is identity operator in the first $2n-2y$ step, it is not hard to infer that
\begin{equation}|\varphi\rangle^{2n-2y}=\sum_{i=0}^{n-1}c_{i}|(n-y)i\bmod\,n\rangle|i\rangle|0\rangle,\end{equation}
\begin{equation}|\varphi\rangle^{2n-2y+2}=\sum_{i=0}^{n-1}c_{i}|(n-y)i+(i+1)\bmod\,n\rangle|i\rangle|1\rangle,\end{equation}
\begin{equation}|\varphi\rangle^{2n}=\sum_{i=0}^{n-1}c_{i}|(n-y)i+(i+1)y\bmod\,n\rangle|i\rangle|0\rangle=|y\rangle(\sum_{i=0}^{n-1}c_i|i\rangle)|1\rangle.\end{equation}

\subsection{Transfer realization of a qubit state on the 2-complete graph}
To transfer the qubit state on the 2-complete graph from $y=0$ to $y=1$, such as $\frac{1}{2}|0\rangle+\frac{\sqrt{3}}{2}|1\rangle$ that can be prepared by using the $U3(\frac{2\pi}{3},0,\frac{\pi}{2})$ gate, we can have the corresponding scheme by the above discussion. The original state can be depicted as $|\varphi\rangle^{0}=|0\rangle(\frac{1}{2}|0\rangle+\frac{\sqrt{3}}{2}|1\rangle)|0\rangle$.
First, let us select the identity operator as the first coin operator, after applying the conditional shift operator, then the state becomes
\begin{equation}|\varphi\rangle^{1}=\frac{1}{2}|0\rangle|00\rangle+\frac{\sqrt{3}}{2}|1\rangle|10\rangle.\end{equation}
Then, choose the second coin operator still to be the identity operator, and after operating the conditional shift operator, we have the state
\begin{equation}|\varphi\rangle^{2}=\frac{1}{2}|0\rangle|00\rangle+\frac{\sqrt{3}}{2}|1\rangle|10\rangle.\end{equation}
After the third  step, the state becomes
\begin{equation}|\varphi\rangle^{3}=\frac{1}{2}|0\rangle|00\rangle+\frac{\sqrt{3}}{2}|0\rangle|10\rangle.\end{equation}
In the fourth step, we select the Pauli $X$ operator as the second coin operator, after using the conditional shift operator, then the state becomes
\begin{equation}|\varphi\rangle^{4}=\frac{1}{2}|1\rangle|01\rangle+\frac{\sqrt{3}}{2}|1\rangle|11\rangle=|1\rangle(\frac{1}{2}|0\rangle+\frac{\sqrt{3}}{2}|1\rangle)|1\rangle.\end{equation}

\begin{figure}[htb]
 \centering
 \subfigure[]{\label{fig:2completepositionibm2}
 \includegraphics[width=3cm]{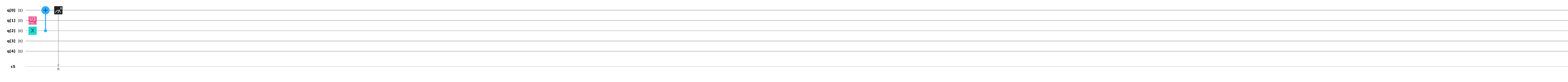}}
 \qquad
 \subfigure[]{\label{fig:2completestateibm2}
 \includegraphics[width=3cm]{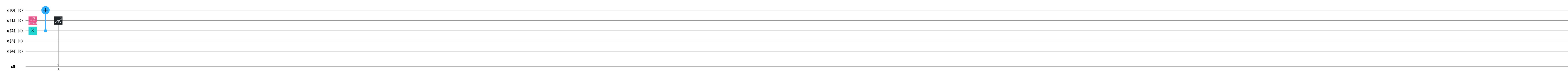}}
 \qquad
 \subfigure[]{\label{fig:2completepositionvigo}
 \includegraphics[width=4cm]{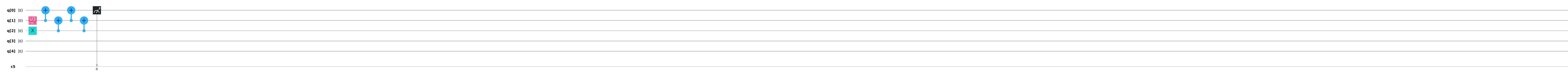}}
 \qquad
 \subfigure[]{\label{fig:2completestatevigo}
 \includegraphics[width=4cm]{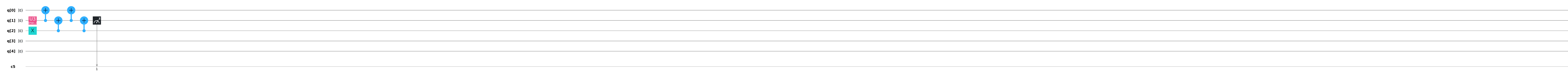}}
 \renewcommand{\figurename}{Figure}
 \caption{Quantum circuit diagram of qubit state transfer on 2-complete graph from $y=0$ to $y=1$. $q[0]$ stands for the position space. $q[1]$ and $q[2]$ stand for 1-st coin space and 2-nd coin space respectively. (a), (c) and (b),(d) stand for the measurements in Z-basis for the position space and 1-st coin space respectively. (a) and (b) are performed on the ``$ibmqx2$" and ``$simulator$". (c) and (d) are performed on the ``$ibmq_{\_}vigo$". Here we use the $U3(\frac{2\pi}{3},0,\frac{\pi}{2})$ gate.}
 \label{fig:2completepositionstate}
 \end{figure}

\begin{table}[htb]
 \centering
  \subtable[]{
  \begin{tabular}{|c|c|c|c|c|}
     \hline
     \multicolumn{1}{|c|}{} & \multicolumn{2}{|c|}{1-st coin state} & \multicolumn{2}{|c|}{position} \\
     \hline
     \diagbox{}{probability}{}           & $|\textbf{0}\rangle$ & $|\textbf{1}\rangle$ &  $|0\rangle$  & $|\textbf{1}\rangle$    \\
     \hline
     ($ibmqx2$) Run 1  & \textbf{0.28845} & \textbf{0.71155} & 0.0188  & \textbf{0.9812}    \\
     \hline
     ($ibmqx2$) Run 2  & \textbf{0.29468} & \textbf{0.70532} & 0.02209 & \textbf{0.97791}     \\
     \hline
     ($ibmqx2$) Run 3  & \textbf{0.29541} & \textbf{0.70459} & 0.02246  & \textbf{0.97754}   \\
     \hline
     ($ibmqx2$) Run 4  & \textbf{0.2948}  & \textbf{0.7052}  & 0.01746  & \textbf{0.98254}   \\
     \hline
     ($ibmqx2$) Run 5  & \textbf{0.29529} & \textbf{0.70471} & 0.01965  & \textbf{0.98035}    \\
     \hline
     ($ibmqx2$) Run 6  & \textbf{0.29968} & \textbf{0.70032} & 0.02161  & \textbf{0.97839}    \\
     \hline
     ($ibmqx2$) Run 7  & \textbf{0.29358} & \textbf{0.70642} & 0.02051  &  \textbf{0.97949}    \\
     \hline
     ($ibmqx2$) Run 8  & \textbf{0.29553} & \textbf{0.70447} & 0.01892  &  \textbf{0.98108}    \\
     \hline
     ($ibmqx2$) Run 9  & \textbf{0.30188} & \textbf{0.69812} & 0.02136  &  \textbf{0.97864}    \\
     \hline
     ($ibmqx2$) Run 10 & \textbf{0.29211} & \textbf{0.70789} & 0.02002  &  \textbf{0.97998}    \\
     \hline
     simulator         & \textbf{0.24341} & \textbf{0.75659} &    0    &   \textbf{1}    \\
     \hline
   \end{tabular}
   \label{Table:results of scheme of 2-complete on ibm2}
  }
 \qquad
  \subtable[]{
  \begin{tabular}{|c|c|c|c|c|}
     \hline
     \multicolumn{1}{|c|}{} & \multicolumn{2}{|c|}{1-st coin state} & \multicolumn{2}{|c|}{position} \\
     \hline
     \diagbox{}{probability}{}           & $|\textbf{0}\rangle$ & $|\textbf{1}\rangle$ &  $|0\rangle$  & $|\textbf{1}\rangle$    \\
     \hline
     ($ibmq_{\_}vigo$) Run 1  & \textbf{0.32227} & \textbf{0.67773} & 0.10681  & \textbf{0.89319}    \\
     \hline
     ($ibmq_{\_}vigo$) Run 2  & \textbf{0.32556} & \textbf{0.67444} & 0.10974 & \textbf{0.89026}     \\
     \hline
     ($ibmq_{\_}vigo$) Run 3  & \textbf{0.31641} & \textbf{0.68359} & 0.1106  & \textbf{0.8894}   \\
     \hline
     ($ibmq_{\_}vigo$) Run 4  & \textbf{0.32886} & \textbf{0.67114} & 0.12256  & \textbf{0.87744}   \\
     \hline
     ($ibmq_{\_}vigo$) Run 5  & \textbf{0.32422} & \textbf{0.67578} & 0.11145  & \textbf{0.88855}    \\
     \hline
     ($ibmq_{\_}vigo$) Run 6  & \textbf{0.32239} & \textbf{0.67761} & 0.11328  & \textbf{0.88672}    \\
     \hline
     ($ibmq_{\_}vigo$) Run 7  & \textbf{0.32043} & \textbf{0.67957} & 0.11609  &  \textbf{0.88391}    \\
     \hline
     ($ibmq_{\_}vigo$) Run 8  & \textbf{0.32312} & \textbf{0.67688} & 0.11072  &  \textbf{0.88928}    \\
     \hline
     ($ibmq_{\_}vigo$) Run 9  & \textbf{0.32007} & \textbf{0.67993} & 0.12036  &  \textbf{0.87964}    \\
     \hline
     ($ibmq_{\_}vigo$) Run 10 & \textbf{0.33044} & \textbf{0.66956} & 0.1145  &  \textbf{0.8855}    \\
     \hline
     theoretical result       & \textbf{0.25}    & \textbf{0.75}    &    0     &   \textbf{1}    \\
     \hline
   \end{tabular}
   \label{Table:results of scheme of 2-complete on vigo}
  }
  \caption{Results of the single qubit state transfer on the 2-complete graph from $y=0$ to $y=1$. (a) and (b) are the results corresponding to ``$ibmqx2$" and ``$ibmq_{\_}vigo$" respectively.}
  \label{Table:results of scheme of 2-complete}
 \end{table}

Indeed, we find that the first three steps in this scheme don't work at all, but they would introduce general noise and cause decoherence in the process of evolution.
Thus we can simplify this scheme which just use one-step quantum walk.
Based on the scheme, we run it on the ``$ibmqx2$" and ``$simulator$" which can be seen in Figure \ref{fig:2completepositionibm2} and Figure \ref{fig:2completestateibm2}.
In addition, we also run the experiment on the ``$ibmq_{\_}vigo$" which designed based on its connectivity. The relevant experimental implementation is shown in Figure \ref{fig:2completepositionvigo} and Figure \ref{fig:2completestatevigo}.
We run experiments 10 times in total. For each experiment, 8192 shots were used that can reduce the statistical error as much as possible.
Table \ref{Table:results of scheme of 2-complete} shows the probability results.

According to Table \ref{Table:results of scheme of 2-complete}, neither the run results on ``$ibmq_{\_}vigo$" nor the results on ``$ibmqx2$" are as close to the theoretical value as ``$simulator$".
During the implementation of this quantum circuit, the decoherence of the given quantum state would increase with the application of many gates in sequence. And then it would produce more noise in the system. Thus, it is hard for us to realize the performance of a quantum circuit with exact accuracy in terms of the current level of the platform.
As for these two quantum chips (``$ibmq_{\_}vigo$" and ``$ibmqx2$"), we now compare their performances on this task by proceeding quantum state tomography. The data is presented in Appendix B. Just like the discussions in Section 3.2, we can get:
\begin{equation}
\rho^{T}=
\left(
  \begin{array}{cc}
    0.2500 & 0.4330 \\
    0.4330 & 0.7500
  \end{array}
\right),
\end{equation}
\begin{equation}
\rho^{E}_{ibmqx2}=
\left(
  \begin{array}{cc}
    0.2951         & 0.4297-0.0006i \\
    0.4297+0.0006i & 0.7049
  \end{array}
\right),
\end{equation}
\begin{equation}
\rho^{E}_{ibmq_{\_}vigo}=
\left(
  \begin{array}{cc}
    0.3234         & 0.4225-0.0238i \\
    0.4225+0.0238i & 0.6766
  \end{array}
\right).
\end{equation}
The fidelity between theoretical and experiment density matrix could be calculated as follows:
\begin{equation}F(\rho^{T},\rho^{E}_{ibmqx2})=(tr(\sqrt{\sqrt{\rho^{T}}\rho^{E}_{ibmqx2}\sqrt{\rho^{T}}}))^{2}=0.9756,\end{equation}
\begin{equation}F(\rho^{T},\rho^{E}_{ibmq_{\_}vigo})=(tr(\sqrt{\sqrt{\rho^{T}}\rho^{E}_{ibmq_{\_}vigo}\sqrt{\rho^{T}}}))^{2}=0.9555.\end{equation}
Thus, the run result on ``$ibmq_{\_}vigo$" is less better than ``$ibmqx2$".
The reason is that the error would increase when we increase the depth of the quantum circuit and the quantum circuit performed on the ``$ibmq_{\_}vigo$" is a little bit more complicated which is limited by its connectivity.
Overall, the results on the quantum machine are good.

\subsection{Transfer realization of two qubits state on the 4-complete graph }
To transfer two qubits state on the 4-complete graph from $y=0$ to $y=1$, such as the Bell state $\frac{|00\rangle+|11\rangle}{\sqrt{2}}$, we can also have the corresponding scheme according to the above discussion in Section 4. The basis state of the position space can be written as $|00\rangle$, $|01\rangle$, $|10\rangle$, and $|11\rangle$. We take this as an example. Thus, the initial state of the whole system could be depicted as
\begin{equation}|\phi\rangle^{0}=|00\rangle(\frac{|00\rangle+|11\rangle}{\sqrt{2}})|00\rangle.\end{equation}
In the first step, we choose $C_{1}=I$, after using the conditional shift operator, then the state becomes
\begin{equation}|\phi\rangle^{1}=\frac{1}{\sqrt{2}}(|00\rangle|00\rangle|00\rangle+|11\rangle|11\rangle|00\rangle).\end{equation}
Furthermore, we could get
\begin{equation}|\phi\rangle^{2}=|\phi\rangle^{1}, \ \ \ \ \ \ |\phi\rangle^{3}=\frac{1}{\sqrt{2}}(|00\rangle|00\rangle|00\rangle+|10\rangle|11\rangle|00\rangle),\end{equation}
\begin{equation}|\phi\rangle^{4}=|\phi\rangle^{3}, \ \ \ \ \ \ |\phi\rangle^{5}=\frac{1}{\sqrt{2}}(|00\rangle|00\rangle|00\rangle+|01\rangle|11\rangle|00\rangle),\end{equation}
\begin{equation}|\phi\rangle^{6}=|\phi\rangle^{5}, \ \ \ \ \ \ |\phi\rangle^{7}=\frac{1}{\sqrt{2}}(|00\rangle|00\rangle|00\rangle+|00\rangle|11\rangle|00\rangle),\end{equation}
in turn.
And in the last step, the coin operator is generalized Pauli $X$ operator (the order is four), so the final state is
\begin{equation}|\phi\rangle^{8}=\frac{1}{\sqrt{2}}(|01\rangle|00\rangle|01\rangle+|01\rangle|11\rangle|01\rangle)=|01\rangle(\frac{|00\rangle+|11\rangle}{\sqrt{2}})|01\rangle.\end{equation}

\begin{figure}[htb]
 \centering
 \subfigure[]{\label{fig:4completeposition}
 \includegraphics[width=7cm]{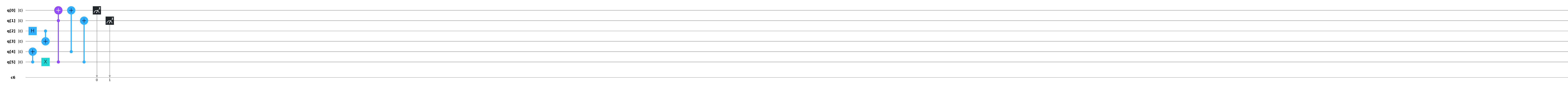}}
 \subfigure[]{\label{fig:4completestate}
 \qquad
 \includegraphics[width=7cm]{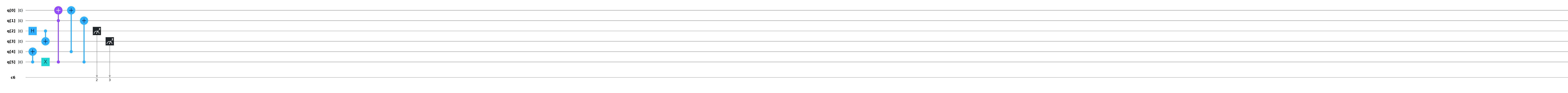}}
 \renewcommand{\figurename}{Figure}
 \caption{Quantum circuit diagram of Bell state transfer on 4-complete graph from $y=0$ to $y=1$. $q[0]$, $q[1]$ stand for the position space. $q[2]$, $q[3]$ and $q[4]$, $q[5]$ stand for 1-st coin space and 2-nd coin space respectively. (a) and (b) stand for the measurements in Z-basis for the position space and 1-st coin space respectively.}
 \label{fig:4completepositionstate}
 \end{figure}

Just like the above discussion in Section 4.1, we observe that the first seven steps do not work at all. For a better experimental realization, we could run it on IBM Quantum Experience platform using one-step coined quantum walk, which can be seen in Figure \ref{fig:4completepositionstate}.
Since this quantum circuit has 6 qubits in total, we use ``$ibmq_{\_}16$" quantum chip to perform the corresponding experiment here.
So we have to design the corresponding quantum circuit that accords with the connectivity of ``$ibmq_{\_}16$" quantum chip \cite{IBM_2017}.
In order to achieve the $Toffoli$ gate $T_{1\bigwedge5,0}$ shown in Figure \ref{fig:4completepositionstate}, we need to realize $C_{4,0}$ and $C_{5,1}$ according to the connectivity of ``$ibmq_{\_}16$" \cite{Nielsen_2002}. They can be described as follows,
\begin{equation}
\begin{split}
C_{5,1}&=C_{5,3}C_{3,1}C_{5,3}C_{3,1}=C_{5,3} \cdot (H\otimes H)C_{1,3}(H\otimes H) \cdot C_{5,3}\cdot(H\otimes H)C_{1,3}(H\otimes H)\\
       &=C_{5,4}C_{4,3}C_{5,4}C_{4,3} \cdot (H\otimes H)C_{1,2}C_{2,3}C_{1,2}C_{2,3}(H\otimes H) \cdot C_{5,4}C_{4,3}C_{5,4}C_{4,3} \cdot (H\otimes H)C_{1,2}C_{2,3}C_{1,2}C_{2,3}(H\otimes H)
\end{split}
\end{equation}
\begin{equation}
\begin{split}
C_{4,0}&=C_{4,3}C_{3,0}C_{4,3}C_{3,0}=C_{4,3} \cdot C_{3,1}C_{1,0}C_{3,1}C_{1,0} \cdot C_{4,3} \cdot C_{3,1}C_{1,0}C_{3,1}C_{1,0}\\
       &=C_{4,3} \cdot (H\otimes H)C_{1,3}(H\otimes H)C_{1,0}(H\otimes H)C_{1,3}(H\otimes H))C_{1,0} \cdot C_{4,3} \cdot (H\otimes H)C_{1,3}(H\otimes H))C_{1,0}(H\otimes H)C_{1,3}(H\otimes H)C_{1,0}\\
       &=C_{4,3} \cdot (H\otimes H)C_{1,2}C_{2,3}C_{1,2}C_{2,3}(H\otimes H)C_{1,0}(H\otimes H)C_{1,2}C_{2,3}C_{1,2}C_{2,3}(H\otimes H))C_{1,0}\\
       &\ \ \cdot C_{4,3} \cdot (H\otimes H)C_{1,2}C_{2,3}C_{1,2}C_{2,3}(H\otimes H)C_{1,0}(H\otimes H)C_{1,2}C_{2,3}C_{1,2}C_{2,3}(H\otimes H)C_{1,0}
\end{split}
\end{equation}

The above two results are shown in Table \ref{Table:results of scheme of 4-complete}. Also, the number of shots for each experiment is 8192. Compared with the simulation result, therefore, the run result on ``$ibmq_{\_}16$" quantum chip is still less accurate.
In fact, the quantum circuit that can be performed on the ``$ibmq_{\_}16$" becomes very complex (it takes about 317 time slots). Thus, it is inevitable to produce noises and decoherence.

\begin{table}[htb]
  \centering
  \begin{tabular}{|c|c|c|c|c|c|c|c|c|}
     \hline
     \multicolumn{1}{|c|}{} & \multicolumn{4}{|c|}{1-st coin state} & \multicolumn{4}{|c|}{position} \\
     \hline
     \diagbox{}{probability}{}         & $|\textbf{00}\rangle$ & $|01\rangle$ &  $|10\rangle$  & $|\textbf{11}\rangle$ &  $|00\rangle$   &  $|\textbf{01}\rangle$   &   $|10\rangle$  &  $|11\rangle$   \\
     \hline
     $simulator$  & \textbf{0.4939} & 0 & 0 &  \textbf{0.5061}  &  0 &  \textbf{1}  & 0 &  0    \\
     \hline
     $ibmq_{\_}16$   & \textbf{0.31299} & 0.29614  & 0.20386 & \textbf{0.18701} &  0.31775  & \textbf{0.28894} &  0.20337 & 0.18994    \\
     \hline
     theoretical values      & \textbf{0.5} & 0  &  0  & \textbf{0.5}  & 0  & \textbf{1}  & 0 &  0    \\
     \hline
   \end{tabular}
  \caption{Results of the Bell state transfer on the 4-complete graph from $y=0$ to $y=1$.}
  \label{Table:results of scheme of 4-complete}
\end{table}

\subsection{Transfer realization of three qubits states on the 8-complete graph }
In this part, we discuss three qubits state transfer on the 8-complete graph from $y=0$ to $y=4$.
Take the GHZ state $\frac{|000\rangle+|111\rangle}{\sqrt{2}}$ as an example. In the same way, we could use the binary representation of vertex. Naturally,  we could establish a one-to-one correspondence between $\{|0\rangle, |1\rangle, \cdots, |7\rangle\}$ and $\{|000\rangle, |001\rangle, \cdots, |111\rangle\}$.
Therefore the original state of this system could be described as
\begin{equation}|\phi\rangle^{0}=|000\rangle(\frac{|000\rangle+|111\rangle}{\sqrt{2}})|000\rangle.\end{equation}
According to the related discussion in Section 4, we can perform the corresponding operations step by step. We have
\begin{equation}|\phi\rangle^{1}=\frac{1}{\sqrt{2}}(|000\rangle|000\rangle|000\rangle+|111\rangle|111\rangle|000\rangle),\ \ \ \ |\phi\rangle^{2}=|\phi\rangle^{1},\end{equation}
\begin{equation}|\phi\rangle^{3}=\frac{1}{\sqrt{2}}(|000\rangle|000\rangle|000\rangle+|110\rangle|111\rangle|000\rangle),\ \ \ \ |\phi\rangle^{4}=|\phi\rangle^{3},\end{equation}
\begin{equation}|\phi\rangle^{5}=\frac{1}{\sqrt{2}}(|000\rangle|000\rangle|000\rangle+|101\rangle|111\rangle|000\rangle),\ \ \ \ |\phi\rangle^{6}=|\phi\rangle^{5},\end{equation}
\begin{equation}|\phi\rangle^{7}=\frac{1}{\sqrt{2}}(|000\rangle|000\rangle|000\rangle+|100\rangle|111\rangle|000\rangle),\ \ \ \ |\phi\rangle^{8}=|\phi\rangle^{7},\end{equation}
\begin{equation}|\phi\rangle^{9}=\frac{1}{\sqrt{2}}(|000\rangle|000\rangle|000\rangle+|011\rangle|111\rangle|000\rangle).\end{equation}
And in the 10-th step, the coin operator is generalized Pauli $X$ operator (the order is eight), the state becomes
\begin{equation}|\phi\rangle^{10}=\frac{1}{\sqrt{2}}(|001\rangle|000\rangle|001\rangle+|100\rangle|111\rangle|001\rangle).\end{equation}
Finally, we obtain
\begin{equation}|\phi\rangle^{16}=|100\rangle(\frac{|000\rangle+|111\rangle}{\sqrt{2}})|001\rangle.\end{equation}
Similarly, we find the 2nd, 4th, 6th and 8th steps do not work at all. So the whole process takes 12 steps.

\begin{figure}[htb]
 \centering
 \subfigure[]{\label{fig:8completeinitial}
 \includegraphics[width=5.5cm]{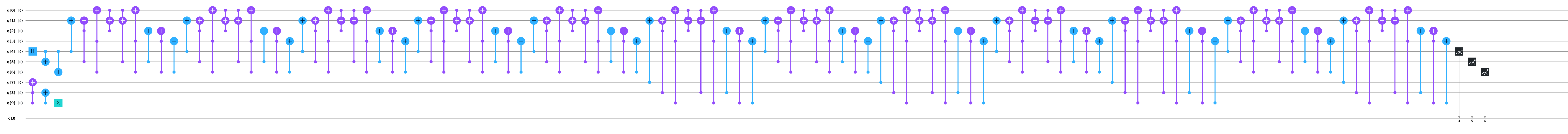}}
 \qquad
 \subfigure[]{\label{fig:completecoin}
 \includegraphics[width=5.5cm]{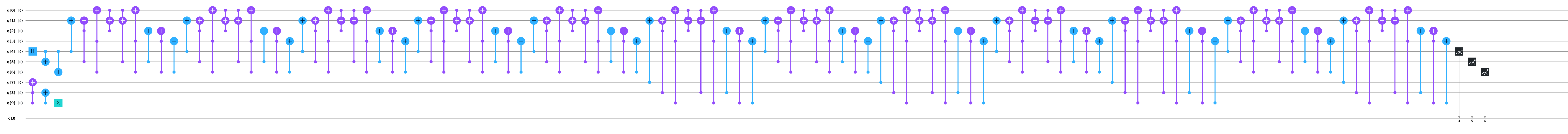}}
 \qquad
 \subfigure[]{\label{fig:8completeS1}
 \includegraphics[width=7cm]{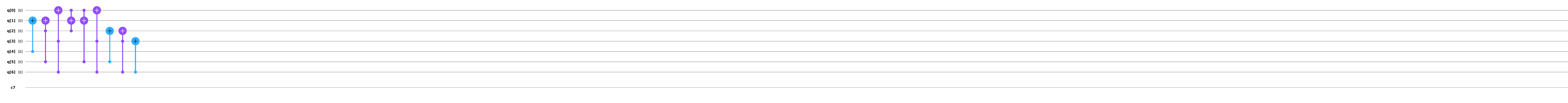}}
 \qquad
 \subfigure[]{\label{fig:8completeS2}
 \includegraphics[width=7cm]{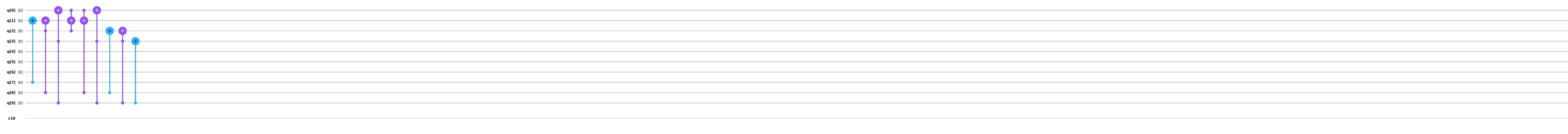}}
 \renewcommand{\figurename}{Figure}
 \caption{Quantum circuit diagram of GHZ state transfer on 8-complete graph from $y=0$ to $y=4$. $q[0]$ stands for the ancillary qubit. $q[1]$, $q[2]$ and $q[3]$ stand for position space. $q[4]$, $q[5]$, $q[6]$ and $q[7]$, $q[8]$, $q[9]$ stand for 1-st coin space and 2-nd coin space respectively. (a) shows the process of preparing GHZ state. (b) is the circuit for generalized Pauli X coin operator. (c) and (d) stand for the conditional shift operator at odd step and even step respectively.}
 \label{fig:8completecomponent}
 \end{figure}

\begin{table}[htb]
  \centering
  \begin{tabular}{|c|c|c|c|c|c|c|c|c|}
     \hline
     \multicolumn{1}{|c|}{} & \multicolumn{2}{|c|}{1-st coin state} & \multicolumn{1}{|c|}{position} \\
     \hline
     \diagbox{}{probability}{}               & $|\textbf{000}\rangle$ & $|\textbf{111}\rangle$ &  $|\textbf{100}\rangle$   \\
     \hline
     simulation 1 (1024)    & \textbf{0.4834} & \textbf{0.5166} & \textbf{1}    \\
     \hline
     simulation 2 (4096)    & \textbf{0.50244} & \textbf{0.49756}  & \textbf{1}    \\
     \hline
     simulation 3 (8192)    & \textbf{0.50085} & \textbf{0.49915}  & \textbf{1}    \\
     \hline
     theoretical values      & \textbf{0.5} & \textbf{0.5}  &  \textbf{1}   \\
     \hline
   \end{tabular}
  \caption{Results of the GHZ state transfer on the 8-complete graph from $y=0$ to $y=4$.}
  \label{Table:results of scheme of 8-complete}
\end{table}

Because the length of quantum circuit performed on the IBM quantum platform is a little deep, we only give the four main components of this quantum circuit here. They are initial state, generalized Pauli X coin operator (the order is 8) and two conditional shift operators (at odd step and even step) shown in Figure \ref{fig:8completecomponent}. This quantum circuit is performed on $simulator$ provided by IBM quantum experience platform. We measure the position space and 1-st coin space in Z-basis to detect whether the GHZ state is transferred from $y=0$ to $y=4$ on the 8-complete graph or not. After running the experiment using 1024, 4096 and 8192 shots respectively, we can get the results in Table \ref{Table:results of scheme of 8-complete}.

According to the statistical result shown in Table \ref{Table:results of scheme of 8-complete}, we can find that the differences between the experimental results (simulation) provided by using 1024, 4096, 8192 shots and the theoretical values are 1.66, 0.244, 0.085 respectively.
Thus, it is clear that we could improve the statistical accuracy by increasing the number of shots. That is also why we all use 8192 shots in the previous sections.

\begin{figure}[htb]
\centering
\includegraphics[width=8cm]{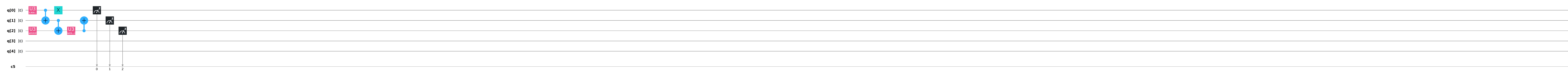}
\caption{Quantum circuit for the generation of W state. $U3(1.9106,0,0)$, $U3(\frac{\pi}{4},0,0)$ and $U3(\frac{\pi}{4},\pi,\pi)$ gates are used.}
\label{fig:Wstate}
\end{figure}

There is an another interesting example, W state, which is given as below \begin{equation}|W\rangle=\frac{|001\rangle+|010\rangle+|100\rangle}{\sqrt{3}}.\end{equation}
To transfer W state on the 8-complete graph, the original state of the whole system is $|\phi\rangle^{0}=|000\rangle(\frac{|001\rangle+|010\rangle+|100\rangle}{\sqrt{3}})|000\rangle$.
According to the protocol discussed in Section 4 and the above discussion about GHZ state, W state can be transferred from $y=0$ to $y=4$ perfectly in the same way.
Because of this similarity, we only give the generation process of W state in detail.
Figure \ref{fig:Wstate} represents the generation of W state.
The gate applied on $q[0]$ is $U3(1.9106,0,0)$ gate (note that $2arccos(\sqrt{\frac{1}{3}})\approx1.9106$)
\begin{equation}
U3(1.9106,0,0)\approx\sqrt{\frac{1}{3}}
\left(
  \begin{array}{cc}
    1         &  -\sqrt{2} \\
    \sqrt{2}  &  1
  \end{array}
\right).
\end{equation}
The two single qubi gates applied on $q[2]$ are $U3(\frac{\pi}{4},0,0)$ and $U3(\frac{\pi}{4},\pi,\pi)$ respectively.
Compared with the creation of W state given in \cite{Chatterjee_2019}, Figure \ref{fig:Wstate} is more simplifier and thus the result is more accurate which can be seen in Table \ref{Table:results of scheme of 8-complete Wstae}.

\begin{table}[htb]
  \centering
  \begin{tabular}{|c|c|c|c|c|c|c|c|c|c|}
     \hline
     \multicolumn{1}{|c|}{}      & \multicolumn{3}{|c|}{1-st coin state} & \multicolumn{1}{|c|}{position} \\
     \hline
     \diagbox{}{probability}{}   & $|\textbf{100}\rangle$ &  $|\textbf{010}\rangle$ & $|\textbf{001}\rangle$ &  $|\textbf{100}\rangle$   \\
     \hline
     simulation 1 (1024)           & \textbf{0.34863}  & \textbf{0.30957} & \textbf{0.34180} & \textbf{1}    \\
     \hline
     simulation 2 (4096)           & \textbf{0.35278}  & \textbf{0.31714} & \textbf{0.33008}  & \textbf{1}    \\
     \hline
     simulation 3 (8192)           & \textbf{0.32666}  & \textbf{0.33838} & \textbf{0.33496}  & \textbf{1}    \\
     \hline
     theoretical values          & \textbf{0.3333\.{3}}  & \textbf{0.3333\.{3}} & \textbf{0.3333\.{3}}  &  \textbf{1}   \\
     \hline
   \end{tabular}
  \caption{Results of the W state transfer on the 8-complete graph from $y=0$ to $y=4$.}
  \label{Table:results of scheme of 8-complete Wstae}
\end{table}

\section{Summary}

Quantum information transfer between different locations in complex quantum  networks is a fundamental and significant task for the mature quantum technology in the future. Perfect state transfer means that information appears at a location and after some time it appears at another location with probability 1. This unbelievable result can never be seen in classical random walks. By using quantum walks with two coins as model, we have given some perfect state transfer schemes on general graphs \cite{Shang_2019}. These schemes can be used as a basic building block to complete more complex quantum communication tasks in scalable quantum computing.

In this paper, we give first experimental realization of high dimensional state transfer on complete graph with two coined quantum walks by implementing the scheme on IBM quantum experience platform. Based on the scheme \cite{Shang_2019}, we prepare the initial state (such as Bell state, GHZ state and W state) and design the corresponding quantum circuit for conditional shift operators and coin operators. In some cases, we might need to introduce ancillary qubits. And then we run it on the quantum devices or simulators through IBM quantum platform.
Also, we perform quantum state tomography to check the accuracy of our protocols.
Compared with experimental realization of teleportation scheme in \cite{Chatterjee_2019}, apart from giving simulator results, we make a further experimental contrast between different quantum chips and present a simpler quantum circuit to generate W state.

From the results obtained by performing experiments on IBM Quantum Experience platform, it is obvious that the accuracy of the simulation results is obviously better than that of the quantum chips (Here, we use ``$ibmq_{\_}5_{\_}yorktown-ibmqx2$", ``$ibmq_{\_}vigo$" and ``$ibmq_{\_}16_{\_}melbourne$" quantum chips).
In fact, IBM Quantum Experience software platform itself has the noises, such as decoherence, depolarizing, general noises and so on. So this phenomenon is not surprising.
Due to the difference of connectivity graphs on different quantum chips, the same task will have different performances on different quantum chips. In order to obtain a much better performance, we could choose an appropriate quantum chip to implement experiment.
In short, our theoretical result \cite{Shang_2019} is verified by the platform in a way.
And the fidelity of our protocols is high.
In addition, we could improve statistical accuracy by increasing the number of shots.

In general, we provide the first experiment implementation of perfect state transfer using two coined quantum walks on IBM Quantum Experience. To be sure, the simulation results on simulator are satisfactory, which are very close to the theoretical results. Also, the experimental results on quantum chips still maintain high fidelity when the corresponding quantum circuits are not complex enough. The optimization and improvement of quantum machines on IBM Quantum Experience are urgent and imperative.
In addition to validating the theoretical protocol\cite{Shang_2019}, our results also broaden the research route in the future. We can verify other useful protocols or design more practical schemes with the aid of IBM Quantum Experience. Since quantum walks are universal quantum computing model, our results may be very useful for other experiment setup.

\section{Appendix}
\subsection{Appendix A: Probability results of single state transfer on the 8-cycle on simulator}

Here, we provide the probability results on ``$simulator$" after measurements in Z-basis, X-basis and Y-basis shown in Table \ref{Table:results of scheme of 8-cycle XYZ simulator}.
As is known to all, the eigenvalue decomposition of the three pauli matrices are:
\begin{equation}Z=|0\rangle\langle0|-|1\rangle\langle1|, X=|+\rangle\langle+|-|-\rangle\langle-|, Y=|\times\rangle\langle\times|-|\div\rangle\langle\div|,\end{equation}
where $|\pm\rangle=\frac{1}{\sqrt{2}}(|0\rangle\pm|1\rangle)$, $|\times\rangle=\frac{1}{\sqrt{2}}(|0\rangle+i|1\rangle)$,
and $|\div\rangle=\frac{1}{\sqrt{2}}(|0\rangle-i|1\rangle)$.
\\
According to the data, we could obtain $<Z>_{simulator}=-0.5043$, $<X>_{simulator}=0.8647$, $<Y>_{simulator}=-0.0060$.

\begin{table}[htb]
  \centering
  \begin{tabular}{|c|c|c|c|c|c|c|c|c|}
     \hline
     probability   &$|0\rangle$ &$|1\rangle$  &  $|+\rangle$  & $|-\rangle$  &  $|\times\rangle$  & $|\div\rangle$    \\
     \hline
     Simulation 1   &   0.25354  &   0.74646  &  0.93237 & 0.06763 & 0.49146  & 0.50854   \\
     \hline
     Simulation 2   &   0.24377  &   0.75623  &  0.93652 & 0.06348 & 0.50746  & 0.49524   \\
     \hline
     Simulation 3   &   0.25085  &   0.74915  &  0.93408 & 0.06592 & 0.4969  & 0.50391   \\
     \hline
     Simulation 4   &   0.24951  &   0.75049  &  0.93506 & 0.06494 & 0.49658  & 0.50342   \\
     \hline
     Simulation 5   &   0.25061  &   0.7939   &  0.93396 & 0.06604 & 0.50476  & 0.49524   \\
     \hline
     Simulation 6   &   0.24048  &   0.75952  &  0.93433 & 0.06567 & 0.49841  & 0.50159  \\
     \hline
     Simulation 7   &   0.24292  &   0.75708  &  0.93396 & 0.06604 & 0.48865  & 0.51135   \\
     \hline
     Simulation 8   &   0.25122  &   0.74878  &  0.92432 & 0.07568 & 0.49988  & 0.50012   \\
     \hline
     Simulation 9   &   0.24280  &   0.75720  &  0.93188 & 0.06812 & 0.49036  & 0.50964   \\
     \hline
     Simulation 10  &   0.25281  &   0.74719  &  0.92700 & 0.07300 & 0.49890  & 0.50110   \\
     \hline
     Mean value     &  0.247851  &  0.752149  &  0.932348 &0.067652 & 0.496985 & 0.503015   \\
     \hline
   \end{tabular}
  \caption{Probability results of the single state transfer on the 8-cycle from $x=0$ to $x=5$ obtained by performing measurement in Z-basis (shown in the second and third columns), X-basis (shown in the fourth and fifth columns) and Y-basis (shown in the sixth and seventh columns) on the 1-st coin state space on ``$simulator$".}
  \label{Table:results of scheme of 8-cycle XYZ simulator}
\end{table}

\subsection{Appendix B: Probability results of single state transfer on the 2-complete graph on two quantum chips}

Here, we present the probability results on ``$ibmqx2$" and ``$ibmq_{\_}vigo$" after measurements in X-basis and Y-basis shown in Table \ref{Table:results of scheme of 4-complete XY ibm2} and Table \ref{Table:results of scheme of 2-complete XY vigo} respectively.
From these two tables, we could obtain
$<X>_{ibmqx2}=0.859498$, $<Y>_{ibmqx2}=-0.001124$, $<X>_{ibmq_{\_}vigo}=0.844998$, $<Y>_{ibmq_{\_}vigo}=-0.047508$.
In addition,
$<Z>_{ibmqx2}=-0.409718$ and  $<Z>_{ibmq_{\_}vigo}=-0.353246$ can be calculated by Table \ref{Table:results of scheme of 2-complete}.

\begin{table}[htb]
  \centering
  \begin{tabular}{|c|c|c|c|c|c|c|c|c|}
     \hline
     probability       & $|+\rangle$ & $|-\rangle$ &  $|\times\rangle$  & $|\div\rangle$    \\
     \hline
     ($ibmqx2$) Run 1  & 0.92676 & 0.07324 & 0.49585  & 0.50415    \\
     \hline
     ($ibmqx2$) Run 2  & 0.92993 & 0.07007 & 0.50244  & 0.49756     \\
     \hline
     ($ibmqx2$) Run 3  & 0.93420 & 0.06580 & 0.50134  & 0.49866   \\
     \hline
     ($ibmqx2$) Run 4  & 0.93384 & 0.06616 & 0.49255  & 0.50745   \\
     \hline
     ($ibmqx2$) Run 5  & 0.92078 & 0.07922 & 0.50085  & 0.49915    \\
     \hline
     ($ibmqx2$) Run 6  & 0.92493 & 0.07507 & 0.49646  & 0.50354    \\
     \hline
     ($ibmqx2$) Run 7  & 0.93433 & 0.06567 & 0.49463  & 0.50537    \\
     \hline
     ($ibmqx2$) Run 8  & 0.92212 & 0.07788 & 0.51099  & 0.48901    \\
     \hline
     ($ibmqx2$) Run 9  & 0.93713 & 0.06287 & 0.49927  & 0.50073    \\
     \hline
     ($ibmqx2$) Run 10 & 0.93347 & 0.06653 & 0.50000  & 0.50000    \\
     \hline
     Mean value        & 0.929749 & 0.070251 & 0.499438  &  0.500562    \\
     \hline
   \end{tabular}
  \caption{Probability results of the single state transfer on the 2-complete graph from $y=0$ to $y=1$ obtained by performing measurement in X-basis (shown in the second and third columns) and Y-basis (shown in the fourth and fifth columns) on the 1-st coin state space on ``$ibmqx2$".}
  \label{Table:results of scheme of 4-complete XY ibm2}
\end{table}

\begin{table}[htb]
  \centering
  \begin{tabular}{|c|c|c|c|c|}
     \hline
     probability              & $|+\rangle$ & $|-\rangle$ &  $|\times\rangle$  & $|\div\rangle$    \\
     \hline
     ($ibmq_{\_}vigo$) Run 1  & 0.92554 & 0.07446 & 0.46106  & 0.53894    \\
     \hline
     ($ibmq_{\_}vigo$) Run 2  & 0.91968 & 0.08032 & 0.45813  & 0.54187     \\
     \hline
     ($ibmq_{\_}vigo$) Run 3  & 0.92029 & 0.07971 & 0.47205  & 0.52795   \\
     \hline
     ($ibmq_{\_}vigo$) Run 4  & 0.91663 & 0.08337 & 0.49255  & 0.50745   \\
     \hline
     ($ibmq_{\_}vigo$) Run 5  & 0.92261 & 0.07739 & 0.48755  & 0.51245    \\
     \hline
     ($ibmq_{\_}vigo$) Run 6  & 0.92383 & 0.07617 & 0.49146  & 0.50854    \\
     \hline
     ($ibmq_{\_}vigo$) Run 7  & 0.92151 & 0.07849 & 0.49146  & 0.50854    \\
     \hline
     ($ibmq_{\_}vigo$) Run 8  & 0.92297 & 0.07703 & 0.4502   & 0.5498    \\
     \hline
     ($ibmq_{\_}vigo$) Run 9  & 0.92798 & 0.07202 & 0.47668  & 0.52332    \\
     \hline
     ($ibmq_{\_}vigo$) Run 10 & 0.92395 & 0.07605 & 0.48132  & 0.51868    \\
     \hline
     Mean value               &0.922499 & 0.077501& 0.476246 & 0.523754    \\
     \hline
   \end{tabular}
  \caption{Probability results of the single state transfer on the 2-complete graph from $y=0$ to $y=1$ obtained by performing measurement in X-basis (shown in the second and third columns) and Y-basis (shown in the fourth and fifth columns) on the 1-st coin state space on ``$ibmq_{\_}vigo$".}
  \label{Table:results of scheme of 2-complete XY vigo}
 \end{table}

\section*{Acknowledgements}
We thanks the support of National Key Research and Development Program of China under grant 2016YFB1000902, National Natural Science Foundation of China (Grant No.61472412, 61872352), Center for Quantum Computing, Peng Cheng Laboratory, and Program for Creative Research Group of National Natural Science Foundation of China (Grant No. 61621003).


\begin{thebibliography}{99}



\bibitem{Shang_2019}
Shang Y., Wang Y., Li M., and Lu R.,
Quantum communication protocols by quantum walks with two coins.
EPL (Europhysics Letters), 124(6), 60009 (2019).

\bibitem{Bose_2003}
Bose S.,
Quantum communication through an unmodulated spin chain.
Phys. Rev. Lett., 91(20), 207901 (2003).

\bibitem{christandl_2004}
Christandl, M., Datta, N., Ekert, A., and Landahl, A. J.,
Perfect state transfer in quantum spin networks.
Phys. Rev. Lett, 92(18), 187902 (2004).

\bibitem{Burgarth_2005}
Burgarth, D., and Bose, S.,
Conclusive and arbitrarily perfect quantum-state transfer using parallel spin-chain channels.
Phys. Rev. A, 71(5), 052315 (2005).

\bibitem{Gong_2007}
Gong, J., and Brumer, P.,
Controlled quantum-state transfer in a spin chain.
Phys. Rev. A, 75(3), 032331 (2007).

\bibitem{Di_2008}
Di Franco, C., Paternostro, M., and Kim, M. S.,
Perfect state transfer on a spin chain without state initialization.
Phys. Rev. Lett., 101(23), 230502 (2008).

\bibitem{Gualdi_2008}
Gualdi, G., Kostak, V., Marzoli, I., and Tombesi, P.,
Perfect state transfer in long-range interacting spin chains.
Phys. Rev. A, 78(2), 022325 (2008).

\bibitem{Petrosyan_2010}
Petrosyan, D., Nikolopoulos, G. M., and  Lambropoulos, P.,
State transfer in static and dynamic spin chains with disorder.
Phys. Rev. A, 81(4), 042307 (2010).

\bibitem{Nikolopoulos_2014}
Nikolopoulos, G. M., and Jex, I. (Eds.),
Quantum state transfer and network engineering.
Heidelberg: Springer (2014).

\bibitem{Hu_2012}
Hu, X., Fan, H., Zhou, D. L., and Liu, W. M.,
Necessary and sufficient conditions for local creation of quantum correlation.
Phys. Rev. A, 85(3), 032102 (2012).

\bibitem{Li_2009}
Li, Z. G., Fei, S. M., Wang, Z. D., and Liu, W. M.,
Evolution equation of entanglement for bipartite systems.
Phys. Rev. A, 79(2), 024303 (2009).

\bibitem{Abliz_2006}
Abliz, A., Gao, H. J., Xie, X. C., Wu, Y. S., and Liu, W. M.,
Entanglement control in an anisotropic two-qubit Heisenberg X Y Z model with external magnetic fields.
Phys. Rev. A, 74(5), 052105 (2006).




\bibitem{Kendon_2011}
Kendon, V. M., and Tamon, C.
Perfect state transfer in quantum walks on graphs.
J Comput. Theor. Nanos, 8(3), 422-433 (2011).

\bibitem{Yalcinkaya_2015}
Yalcinkaya, I., and Gedik, Z.,
Qubit state transfer via discrete-time quantum walks.
J. Phys A-Math Theor, 48(22), 225302 (2015).

\bibitem{Stefanak_2016}
Stefanak, M., and Skoupy, S.,
Perfect state transfer by means of discrete-time quantum walk search algorithms on highly symmetric graphs.
Phys. Rev. A, 94(2), 022301 (2016).

\bibitem{Nitsche_2016}
Nitsche, T., Elster, F., Novotny, J., G¨¢bris, A., Jex, I., Barkhofen, S., and Silberhorn, C., Quantum walks with dynamical control: graph engineering, initial state preparation and state transfer.
New J. Phys., 18(6), 063017 (2016).




\bibitem{Zhang_2005}
Zhang, J., Long, G. L., Zhang, W., Deng, Z., Liu, W., and Lu, Z.,
Simulation of Heisenberg X Y interactions and realization of a perfect state transfer in spin chains using liquid nuclear magnetic resonance.
Phys. Rev. A, 72(1), 012331 (2005).


\bibitem{Bellec_2012}
Bellec, M., Nikolopoulos, G. M., and Tzortzakis, S.,
Faithful communication Hamiltonian in photonic lattices.
Optics letters, 37(21), 4504-4506 (2012).

\bibitem{Perez-Leija_2013}
Perez-Leija, A., Keil, R., Kay, A., Moya-Cessa, H., Nolte, S., Kwek, L. C., Rodriguez-Lara, B. M., Szameit, A., and Christodoulides, D. N.,
Coherent quantum transport in photonic lattices.
Phys. Rev. A, 87(1), 012309 (2013).

\bibitem{Chapman_2016}
Chapman, R. J., Santandrea, M., Huang, Z., Corrielli, G., Crespi, A., Yung, M. H., Osellame, R., and Peruzzo, A.,
Experimental perfect state transfer of an entangled photonic qubit.
Nature communications, 7, 11339 (2016).

\bibitem{Tang_2018}
Tang, H., Di Franco, C., Shi, Z. Y., He, T. S., Feng, Z., Gao, J., Sun, K., Li, Z. M., Jiao, Z. Q., Wang, T. Y., Kim, M. S., and Jin, X. M.,
Experimental quantum fast hitting on hexagonal graphs.
Nature Photonics, 12(12), 754 (2018).




\bibitem{IBM_2017}
IBM Quantum Experience. https://www.research.ibm.com/ibm-q/

\bibitem{Alsina_2016}
Alsina, D., and Latorre, J. I.,
Experimental test of Mermin inequalities on a five-qubit quantum computer.
Phys. Rev. A 94, 012314 (2016).

\bibitem{Devitt_2016}
Devitt, S. J.,
Performing quantum computing experiments in the cloud.
Phys. Rev. A 94, 032329 (2016).

\bibitem{Behera_2017}
Behera, B. K., Banerjee, A., and Panigrahi, P. K.,
Experimental realization of quantum cheque using a five-qubit quantum computer.
Quantum Inf. Process., 16(12), 312 (2017).

\bibitem{Hebenstreit_2017}
Hebenstreit, M.,  Alsina, D., Latorre, J. I. and Kraus, B.,
Compressed quantum computation using a remote five-qubit quantum computer.
Phys. Rev. A 95, 052339 (2017).

\bibitem{Vishnu_2018}
Vishnu, P. K., Joy, D., Behera, B. K., and Panigrahi, P. K.,
Experimental demonstration of non-local controlled-unitary quantum gates using a five-qubit quantum computer.
Quantum Inf. Process., 17(10), 274 (2018).

\bibitem{Chatterjee_2019}
Chatterjee, Y., Devrari, V., Behera, B. K., and Panigrahi, P. K.,
Experimental realization of quantum teleportation using coined quantum walks.
arXiv preprint arXiv:1908.01348, (2019).






\bibitem{Brun_2003}
Brun, T. A., Carteret, H. A., and Ambainis, A.,
Quantum walks driven by many coins.
Phys. Rev. A 67(5), 052317 (2003).

\bibitem{Nielsen_2002}
Nielsen, M. A., and Chuang, I. L.,
Quantum computation and quantum information.
(2002).

\bibitem{Wang_2017}
Shang Y., Wang Y., Xue P.,
Generalized teleportation by quantum walks.
Quantum Inf. Process. 16, 221 (2017)

\bibitem{Paris_2004}
Paris, M., and Rehacek, J. (Eds.).
Quantum state estimation (Vol. 649).
Springer Science and Business Media (2004).



\end{thebibliography}
\end{document}